\documentclass[preprint, aps, pre, eqsecnum,amsmath, amssymb]{revtex4}              
\usepackage[final]{graphicx} 
\begin{document}

\title{
Periodically forced ferrofluid pendulum: effect of polydispersity 
}

\author{
A.~Leschhorn, M.~L\"ucke  
}
\affiliation{
Theoretische Physik, 
Universit\"{a}t des Saarlandes, D-66041~Saarbr\"{u}cken, Germany\\}

\date{\today}

\begin{abstract}

We investigate a torsional pendulum containing a ferrofluid  
that is forced periodically to undergo small-amplitude oscillations.   
A homogeneous magnetic field is applied perpendicular to the pendulum axis. 
We give an analytical formula for the ferrofluid-induced ``selfenergy'' in
the pendulum's dynamic response function for 
monodisperse as well as for polydisperse ferrofluids.

\end{abstract}

\maketitle                                                                      
 %\tableofcontents

 % skip 2/6inch
\vskip2pc

\section{Introduction}

Real ferrofluids \cite{BIB:rosen} contain magnetic particles of different sizes 
\cite{BIB:vert}. This polydispersity strongly influences the macroscopic 
magnetic properties of the ferrofluid. 
We investigate here the effect of polydispersity on the dynamic response of a 
ferrofluid pendulum. 

\begin{figure}[htp] 
\centerline{\includegraphics[width=12cm,angle=0]{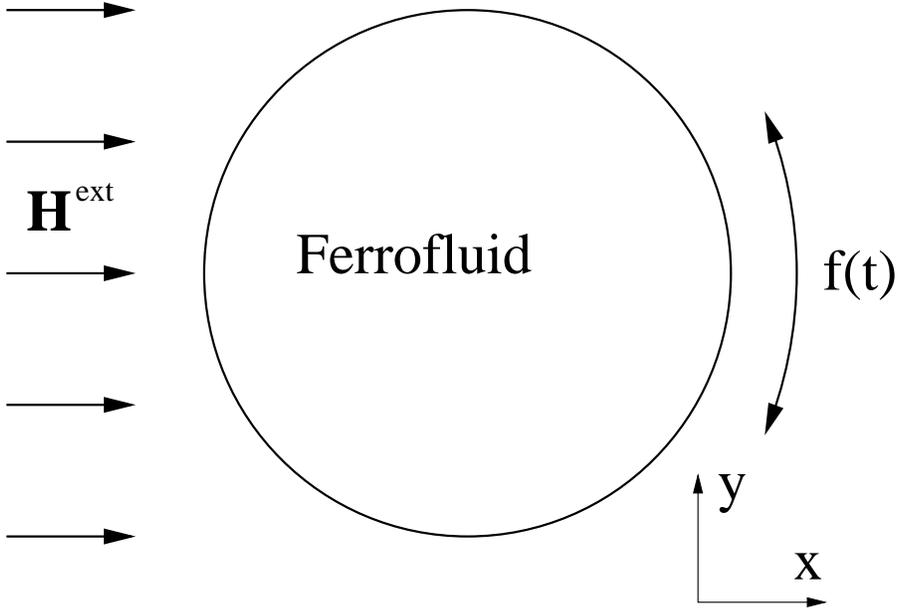}}
\caption{Schematic plot of the system}
\label{FIG:sys}
\end{figure}

A torsional pendulum containing a ferrofluid is forced periodically in   
a homogeneous magnetic field ${\bf H}_{ext}=H_{ext} {\bf e}_x$ that is applied 
perpendicular to the pendulum axis ${\bf e}_z$ (see fig. \ref{FIG:sys}). 
Such a ferrofluid pendulum is used for measuring the rotational viscosity 
\cite{BIB:tp}. 
The cylindrical ferrofluid container is here of sufficiently large length to be 
approximated as an infinite long cylinder. We consider rigid-body rotation of
the ferrofluid with the time dependent angular velocity 
${\bf \Omega} = \dot \varphi{\bf e}_z$ as can be realized with the set-up of 
\cite{BIB:tp}. The fields ${\bf H}$ and ${\bf M}$ inside the 
cylinder are spatially homogeneous and oscillating in time.

\section{Equations} 

First, the Maxwell equations demand that the fields ${\bf H}$ and ${\bf M}$
within the ferrofluid are related to each other via 
\begin{equation} \label{EQ:max} 
{\bf H} + N {\bf M} = {\bf H}^{ext}
\end{equation}
with $N=1/2$ for the infinitely long cylinder.
Then we have the torque balance  
\begin{equation} \label{EQ:tb} 
\ddot \varphi 
= 
- \omega_0^2 \varphi 
- \Gamma_0 \dot \varphi 
- \frac{T}{\Theta} 
+ f(t) 
\end{equation} 
with the eigenfrequency $\omega_0$ and the damping rate 
$\Gamma_0$ of the
pendulum without field and the total moment of inertia $\Theta$. The magnetic 
torque reads 
\begin{equation} \label{EQ:magtorque} 
T 
=  
- \mu_0 \int dV ({\bf M} \times {\bf H})_z 
= 
- \mu_0 V ({\bf M} \times {\bf H}^{ext})_z \, ,
\end{equation} 
and $f(t)$ is the external mechanical forcing. 

Finally, we need an equation describing the magnetization dynamics. 
Here, we consider the polydisperse ferrofluid as a mixture of ideal
monodisperse paramagnetic fluids. Then the resulting magnetization is given by 
${\bf M}=\sum {\bf M}_j$, where ${\bf M}_j$ denotes the magnetization of the
particles with diameter $d_j$. We assume that each ${\bf M}_j$ obeys a 
simple Debye relaxation dynamics described by  
\begin{equation} \label{EQ:maggl} 
d_t {\bf M}_j 
- {\bf \Omega} \times {\bf M}_j  
= 
- \frac{1}{\tau_j} [{\bf M}_j - {\bf M}_j^{eq} ({\bf H})] 
\end{equation} 
We take the equilibrium magnetization to be given by a Langevin function 
\begin{equation} \label{EQ:meq} 
{\bf M}_j^{eq} ({\bf H}) 
= 
\chi_j (H) {\bf H} 
= 
w_j {\cal L}\left( \frac{\mu_0 \pi M_{mat}}{6k_BT} d_j^3 H  \right) 
\, \frac{{\bf H}}{H}
\end{equation} 
with the saturation magnetization of the material $M_{mat}$ and the
magnetization distribution $w_j (d_j)$. 
Note that the magnetization equations (\ref{EQ:maggl}) for the different 
particle sizes are
coupled by the internal field ${\bf H}={\bf H}^{ext}- N {\bf M}$. 
As relaxation rate we combine Brownian and N\'eel relaxation 
$
\frac{1}{\tau_j} 
= 
\frac{1}{\tau_B^j} + \frac{1}{\tau_N^j}
$.
The relaxation times depend on the particle size by 
$\tau_B^j=\frac{\pi \eta}{2k_BT} (d_j+2s)^3$ and 
$\tau_N^j=f_0^{-1} \exp\left(\frac{\pi Kd_j^3}{6k_BT}\right)$ with $\eta$ the 
viscosity, $s$ the thickness of the nonmagnetic particle layer, and $K$ the
anisotropy constant. 

Altogether we use the following system of equations: 
\begin{eqnarray} \label{EQ:GLS}
\dot \varphi 
&=& 
\Omega 
\\ 
\dot \Omega 
&=& 
- \omega_0^2 
\varphi 
- \Gamma_0 
\Omega 
- \mu_0 \frac{V}{\Theta} H^{ext} M_y   
+ f(t) 
\\  
\dot M_x^j  
&=&  
- \Omega M_y^j
- \frac{1}{\tau_j} \left[ M_x^j - \chi_j(H) (H^{ext} - N M_x )\right]  
\\ 
\dot M_y^j 
&=&  
\Omega M_x^j  
- \frac{1}{\tau_j} M_y^j
- \frac{1}{\tau_j} N\chi_j(H) M_y \, .   
\end{eqnarray}

\section{Linear response analysis}

For the equilibrium situation of the unforced pendulum at rest that we denote
in the following by an index 0 one has 
$\varphi_0 = \Omega_0 =M_y^{j0} = 0$ and $M_x^{j0}=M^j_{eq}(H_0)$.
Furthermore, $M_0 = \sum M^j_{eq}(H_0)$ with $H_0$ solving the equation 
$H_0 = H^{ext} - N M_0(H_0)$. 

External forcing with small $|f|$ leads to small deviations of 
$\varphi$, of $\Omega$, and of 
$\delta {\bf H} = {\bf H} - {\bf H}_0 = -N({\bf M} - {\bf M}_0)
= - N\delta {\bf M}/2$ 
from the above described equilibrium state. 
We expand each $\chi_j(H)$ up to linear order in $\delta {\bf H}$
\begin{eqnarray}    
\chi_j (|{\bf H}_0 + \delta {\bf H}|) 
&=& 
\chi_{j0} 
- \chi_{j0}' N \delta M_x 
+ {\cal O} (\delta {\bf H})^2 \,. 
\end{eqnarray} 
Here, $\chi_{j0} = \chi_j (H_0)$ and $\chi_{j0}'$ is the derivative of 
$\chi_{j0}$. Then we get the linearized equations 
\begin{eqnarray} 
\dot \varphi 
&=& 
\Omega 
\label{EQ:LGLS1} \\ 
\dot \Omega 
&=& 
- \omega_0^2 \varphi 
- \Gamma_0 \Omega 
- \kappa y 
+ f(t)  
\\   
\dot x_j  
&=& 
- \frac{1}{\tau_j} x_j  
- \frac{1}{\tau_j} N
(\chi_{j0} + \chi_{j0}' H_0) x 
\\ 
\dot y_j  
&=& 
\Omega x_j^0 
- \frac{1}{\tau_j} y_j 
- \frac{1}{\tau_j} N\chi_{j0} y \,. 
\label{EQ:LGLS4}  
\end{eqnarray}
We have intoduced the abbreviations $x_j = \delta M_x/M_0, \,
x_j^0 = M_x^{j,0}/M_0, \, y_j = \delta M_y/M_0$ and 
$x = \sum_j x_j, \, y = \sum_j y_j$. The strength of the coupling constant
between the mechanical degrees of freedom $\varphi, \Omega$ and the magnetic
ones is $\kappa = \mu_0 H^{ext} M_0 V/\Theta$ .   
 
For periodic forcing $f(t)=\hat f e^{-i\omega t}$ we look for solutions in the
form  
\begin{eqnarray} \label{EQ:ansatz} 
\left( 
\begin{array}{c} 
\varphi(t) \\ \Omega(t) \\ x_j(t) \\ y_j(t) 
\end{array} 
\right) 
&=& 
\left( 
\begin{array}{c} 
\hat \varphi \\ \hat \Omega \\ \hat x_j \\ \hat y_j 
\end{array} 
\right) 
\, e^{-i\omega t} \,. 
\end{eqnarray} 
Inserting the ansatz (\ref{EQ:ansatz}) into the linearized 
equations (\ref{EQ:LGLS1}) --(\ref{EQ:LGLS4}) yields 
\begin{eqnarray} \label{EQ:solut}
\hat \Omega 
&=& 
-i\omega \hat \varphi 
\\ 
\hat x 
&=& 
0 
\, = \, 
\hat x_j
\\ 
\hat y_j 
&=& 
- \left[ 
\frac{i\omega \tau_j}{1-i\omega \tau_j} x_j^0 
- \frac{N \chi_{j0}}{1-i\omega \tau_j}
\frac{\omega}{\kappa} \Sigma
\right] 
\hat \varphi
\\ 
\hat y 
&=& 
- \frac{\omega}{\kappa} \Sigma 
\hat \varphi 
\end{eqnarray}
and 
\begin{equation}
\hat \varphi = G \hat f =
\left[ \omega_0^2 - \omega^2 - i\omega \Gamma_0 - \omega \Sigma 
\right]^{-1} \, \hat f \, .
\end{equation}
The ferrofluid-induced ´selfenergy´ $\Sigma(\omega)$ in the expression for the 
dynamical response function $G(\omega)$ of the torsional pendulum is
\begin{eqnarray} \label{EQ:Sigma_poly} 
\Sigma(\omega) 
&=& 
i \kappa \left( 
1 + N \sum_j \frac{\chi_{j0}}{1 - i\omega \tau_j} 
\right)^{-1}  
\sum_j \frac{\tau_j x_j^0}{1 - i\omega \tau_j} \, . 
\end{eqnarray} 
Its imaginary part changes the damping rate $\Gamma_0$ of the pendulum for
$\kappa=0$, i.e., in zero field. The real part shifts the resonance frequency
of the pendulum. In the special case of a monodisperse ferrofluid on has 
\begin{eqnarray} \label{EQ:Sigma_mono} 
\Sigma(\omega)  
&=& 
\frac{i\kappa \tau}{1 - i\omega \tau + N \chi_0} 
\end{eqnarray}

\section{Results} 

\begin{figure}[htp] 
\centerline{\includegraphics[width=12cm,angle=0]{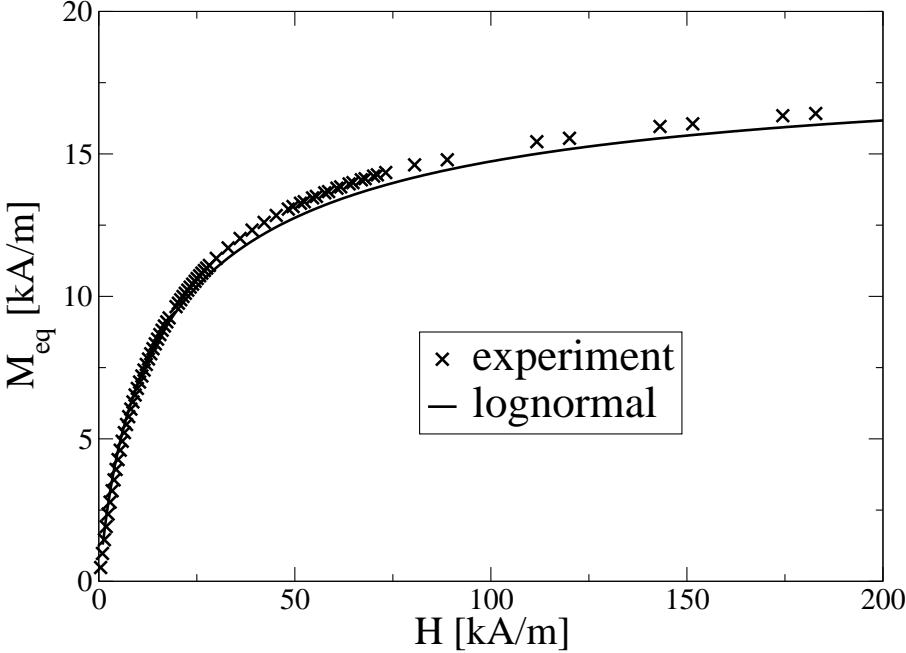}}
\caption{x: Experimental equilibrium magnetization $M_{eq}(H)$ used as input for
the monodisperse calculations;
full line: fit with lognormal contribution.}
\label{FIG:meq}
\end{figure}

\begin{figure} 
\centerline{\includegraphics[width=12cm,angle=0]{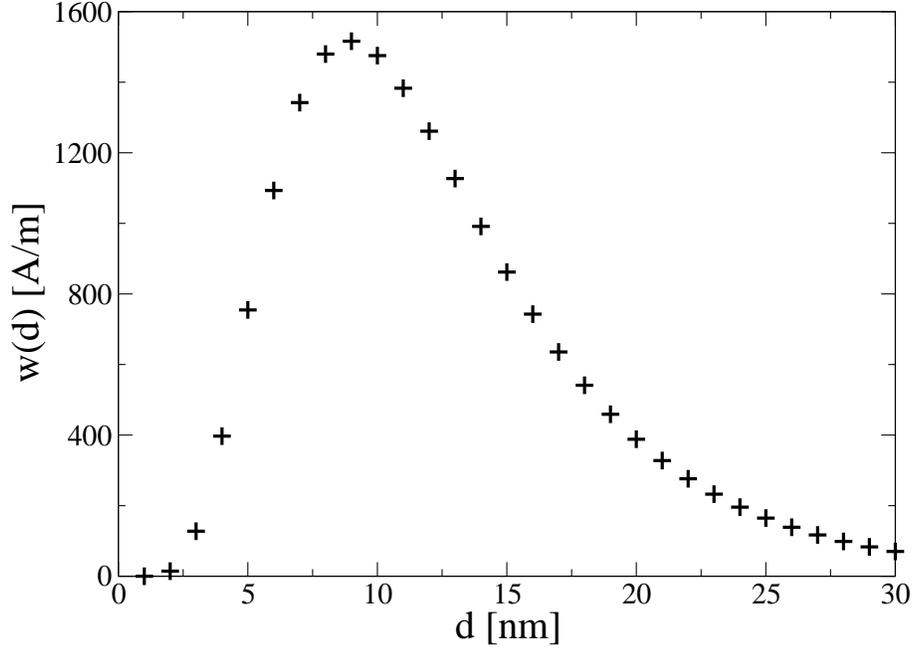}}   
\caption{lognormal contribution $w(d_i)$ ($d_1=1nm \dots d_{30}=30nm$) 
used as input for the polydisperse calculations.}
\label{FIG:maggew}
\end{figure}

We evaluated the linear response function 
$G(\omega)=\hat \varphi(\omega)/\hat f$ of the
pendulum's angular deviation amplitude $\hat \varphi(\omega)$ to the applied 
forcing amplitude $\hat f$ and the 
selfenergy $\Sigma(\omega)$ for some experimental parameters from \cite{BIB:tp}: 
$\omega_0/2\pi = 32.7 Hz$, $\Gamma_0 = 0.178 Hz$, $V/\Theta = 20 m/kg$. 
The cylinder is filled with the ferrofluid APG 933 of FERROTEC. 
Therefore, we used in equation (\ref{EQ:Sigma_mono}) an experimental 
$\tau =0.6ms$ and the experimental $M_{eq}(H)$ shown in fig. \ref{FIG:meq}.
These monodisperse results were compared with the expression
(\ref{EQ:Sigma_poly}) for the polydisperse case for the typical parameter 
values $M_{mat}=456kA/m$, 
$\eta = 0.5 Pa\cdot s$, $s=2nm$, $K=44kJ/m^3$ and $f_0=10^9 Hz$. 
The contributions $w(d_j)$  that enter into the formulas (\ref{EQ:meq}) for the 
susceptibilities $\chi_j$ are given by a lognormal distribution
\cite{BIB:vert}: 
\begin{eqnarray} 
w(d_j) 
= 
M_{sat} \frac{g(d_j) d_j}{\sum_{k=1}^{30} g(d_k) d_k} 
& {\rm with} & 
g(d_j) 
= 
\frac{1}{\sqrt{2\pi} d_j \ln{\sigma}}
\exp{\left(-\frac{\ln^2{(d_j/d_0)}}{2\ln^2{\sigma}}\right)}
\end{eqnarray} 
Fitting the experimental $M_{eq}(H)$ with a sum of Langevin functions 
(\ref{EQ:meq}) yields $M_{sat}=18149A/m$, $d_0=7nm$ and $\sigma = 1.47$
(see fig. \ref{FIG:meq}). 
We used here 30 different particle sizes from $d_1=1nm$ to $d_{30}=30nm$  
(see fig. \ref{FIG:maggew}).

\begin{figure} 
\centerline{\includegraphics[width=12cm,angle=0]{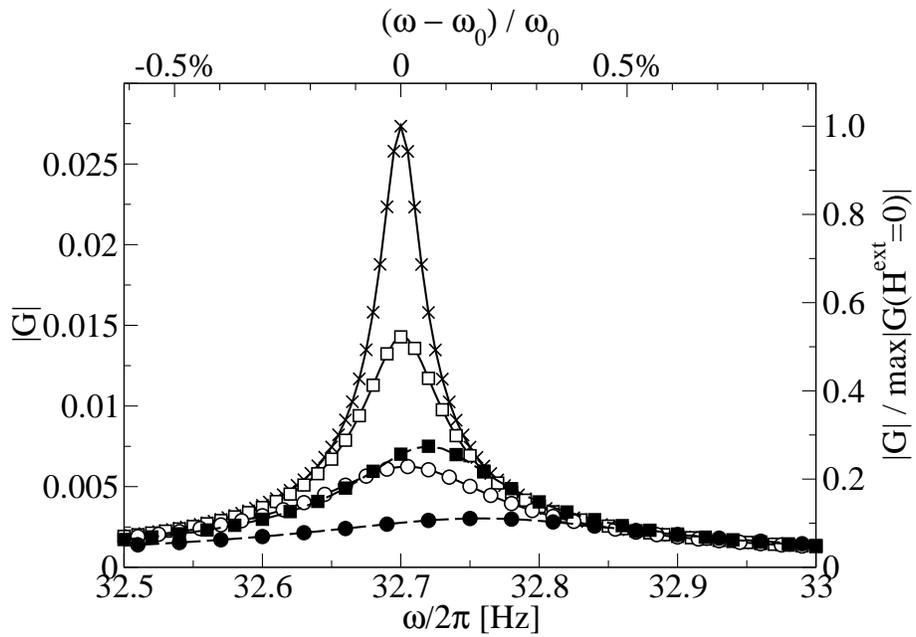}}   
\caption{$|G|$ near the resonance $\omega_0$; 
x $H^{ext} = 0 kA/m$, squares $H^{ext} = 5 kA/m$, circles $H^{ext} = 10 kA/m$; 
filled symbols: polydisperse.}
\label{FIG:absG}
\end{figure}

\begin{figure} 
\centerline{\includegraphics[width=12cm,angle=0]{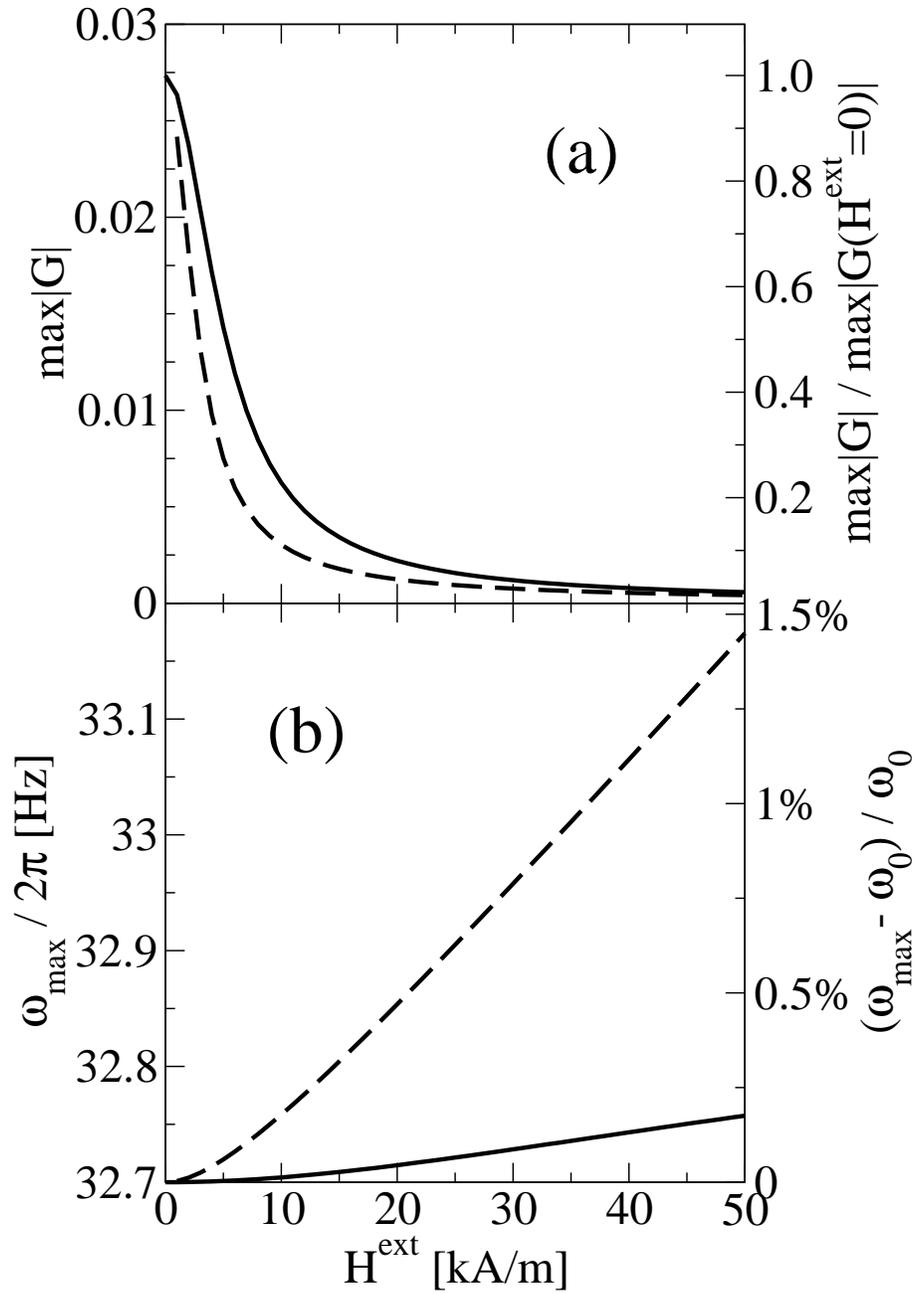}}   
\caption{Maximum value $max|G|$ (a) and peak position $\omega_{max}$ (b)    
as a function of external field $H^{ext}$; 
full line monodisperse, dashed line polydisperse.}
\label{FIG:max}
\end{figure}

The calculations show the additional damping rate caused by the interaction
between ferrofluid and external field. An increasing magnetic field leads to
smaller amplitudes; in polydisperse ferrofluids the amplitude decreases faster
[fig. \ref{FIG:absG} and \ref{FIG:max} (a)]. 
Furthermore, one can see a shift of the peak position to higher frequencies
$\omega_{max}$, which is stronger in polydisperse ferrofluids 
[fig. \ref{FIG:absG} and \ref{FIG:max} (b)].

\begin{acknowledgments}
This work was supported by DFG (SFB 277) and by INTAS(03-51-6064).
\end{acknowledgments}

%%%%%%%%%%%%%%%%%%%%%%%%%%%%%%%%%%%%%%%%%%%%%%%%%%%%%%

\vspace{2cm}

%%%%%%%%%%%%%%%%%%%%%%%%%%%%%%%%%%%%%%%%%%%%%%%%%%%%%

%-------------------------Fig. 1 ---------------------------
%------------------------------------------------------

%-------------------------Fig. 2 ---------------------------
%------------------------------------------------------

%-------------------------Fig. 3 ---------------------------
%----------------------------------------------------------------

%-------------------------Fig. 4 ---------------------------
%----------------------------------------------------------------

%-------------------------Fig. 5 ---------------------------
%----------------------------------------------------------------

%-------------------------Fig. 6 --------------------------- 
\begin{figure} 
\centerline{\includegraphics[width=12cm,angle=0]{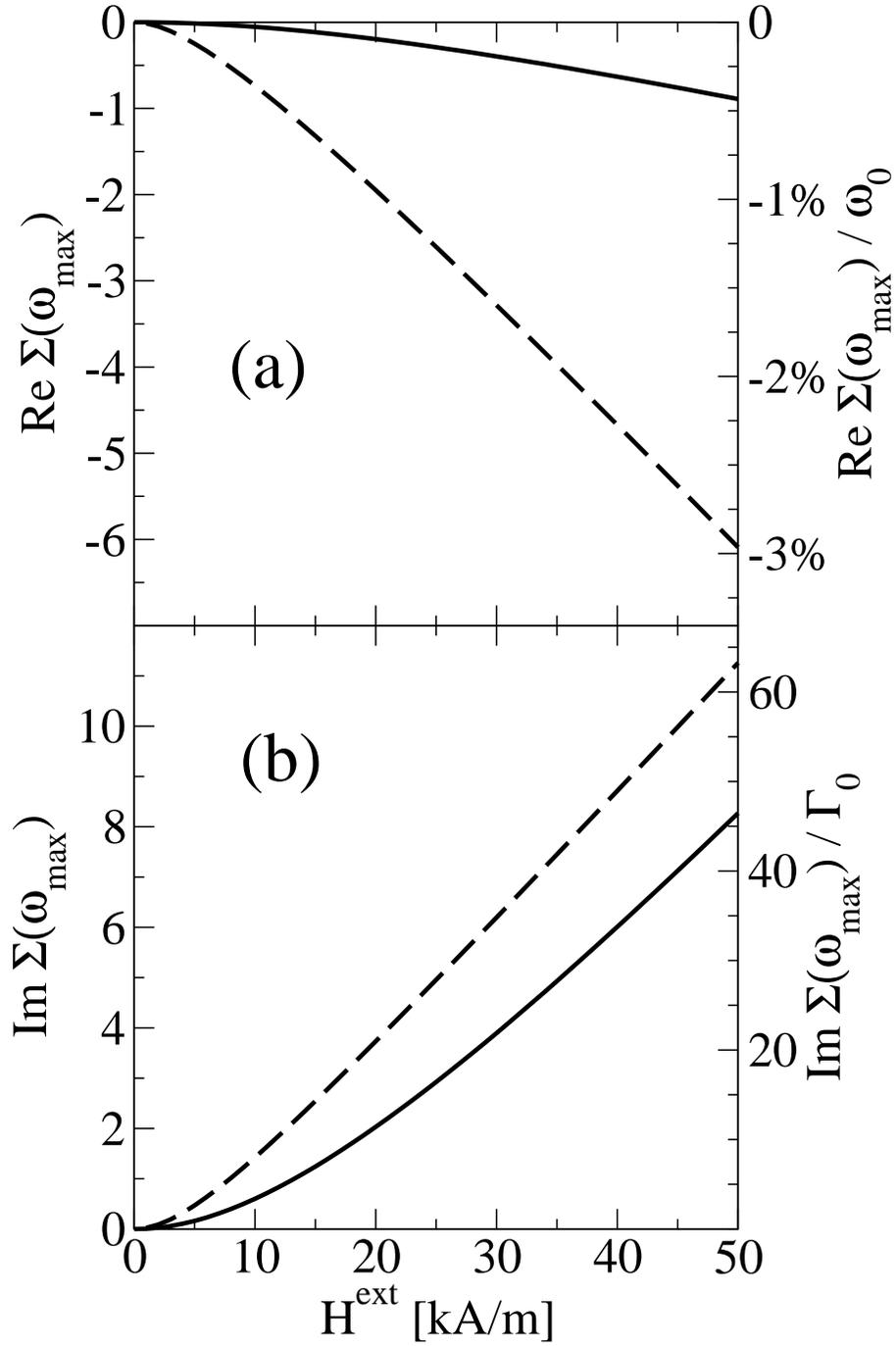}}   
\caption{$Re(\Sigma)$ (a) and $Im(\Sigma)$ (b) at $\omega = \omega_{max}$; 
full line monodisperse, dashed line polydisperse. }
\label{FIG:sigma}
\end{figure} 
%---------------------------------------------------------------- 

%%%%%%%%%%%%%%%%%%%%%%%%%%%%%%%%%%%%%%%%%%%%%%%%%%%%%%%


\begin{thebibliography}{99} 

%-------------------------------------------------------------------------------
\bibitem{BIB:rosen} 
R.~E.~Rosensweig, 
{\it Ferrohydrodynamics}, 
Cambridge University Press, Cambridge (1985). 

%-------------------------------------------------------------------------------
\bibitem{BIB:vert} 
J.~Embs, H.~W.~M\"uller, C.~E.~Krill, F.~Meyer, H.~Natter, B.~M\"uller,
S.~Wiegand, M.~L\"ucke, R.~Hempelmann, K.~Knorr, 
%{\it Particle size analysis of ferrofluids}, 
Magnetohydrodynamics {\bf 37}, 222 (2001).


%-------------------------------------------------------------------------------
\bibitem{BIB:tp}  
J.~Embs, H.~W.~M\"uller, M.~L\"ucke and K.~Knorr, 
%{\it Shear free measurement of the rotational viscosity of ferrofluids with a
% forced torsional pendulum}, 
Magnetohydrodynamics {\bf 36}, 387 (2000);
J.~Embs, H.~W.~M\"uller, C.~Wagner, K.~Knorr and M.~L\"ucke, 
%{\it Measuring the rotational viscosity of ferrofluids without shear flow}, 
Phys. Rev. E {\bf 61}, R2196 (2000). 


\end{thebibliography}
\end{document}